\documentclass[12pt]{article}
\usepackage{amsfonts}
\usepackage{epsfig}
\usepackage{graphicx}
\usepackage{latexsym}    
\usepackage{amsmath}
\usepackage{amssymb}

\newtheorem{lemma}{Lemma}

\newtheorem{theor}{\large\bf Theorem}
\newcommand{\qed}{{\hfill $\Box$}}

\newcommand{\resetequ}{\setcounter{equation}{0}}

\begin{document}
\title{Anderson localization for a supersymmetric sigma model}

\author{M. Disertori$^a$%
\footnote{e-mail: margherita.disertori@univ-rouen.fr},
T. Spencer$^b$ \\
$a$) Laboratoire de Math\'ematiques Rapha\"el Salem, UMR CNRS 6085\\
Universit\'e de Rouen, 76801, France\\
$b$) Institute for Advanced Study, Einstein Drive, \\
Princeton, NJ 08540,  USA\\
}
 \maketitle
\begin{abstract}
We study a lattice sigma model which is expected to reflect the
Anderson localization and delocalization transition for real 
symmetric band matrices in 3D. In this statistical mechanics model, the
field takes values in a supermanifold based on the hyperbolic plane.
The existence of a diffusive phase in 3 dimensions was proved in \cite{DSZ}
for low temperatures. 
Here we prove localization  at high temperatures
for any dimension $d\geq 1$. 
Our analysis uses Ward identities coming from internal supersymmetry.
\end{abstract}

\section{Introduction}

It is well known that the study of localization properties in 
a disordered material can be translated to the study of correlation
functions in a lattice field theory,  
with an internal hyperbolic supersymmetry (SUSY), 
\cite{sw,efetov-Adv,efetov-book, wegner}.
In the physics literature one usually assumes the sigma model 
approximation, which is believed to
capture the essential features of the energy correlations and
transport properties of the underlying quantum system.

The SUSY field theories  which are equivalent to the Anderson 
tight-binding model and random band 
matrices are  difficult to analyze with mathematical
rigor in more than one dimension. In this context, Zirnbauer
introduced a  lattice field model which may be thought
of as a simplified version of one of Efetov's nonlinear sigma models
\cite{toymodel,migdal-kad}. In Zirnbauer's  sigma model the 
field takes values in a target space  $H^{(2|2)}$ which is a 
supermanifold extension of the hyperbolic plane.  The model is 
expected to reflect the spectral properties of random band matrices,
such as  localization and diffusion, in any dimension.
In  \cite{toymodel} localization was established in one
dimensional chain by analyzing the transfer matrix.
We refer to  \cite{DSZ} for a historical 
introduction and motivations.

More recently  the existence of a `diffusive' phase at low
temperatures ($\beta $ large) has been proved for the $H^{(2|2)}$ 
model in  three or more dimensions, see \cite{DSZ}. For $\beta $ small, 
a localized phase was expected. 
However, unlike  conventional  statistical mechanics models, 
the $H^{(2|2)}$ model has a  noncompact hyperbolic symmetry
and so high temperature expansions cannot be done in the usual way. 
In fact, it is known that the bosonic hyperbolic sigma model in 3D has no
localized phase because its effective action is convex for all $\beta >0$. 
On the other hand, numerical simulations \cite{td} indicated that the SUSY 
hyperbolic sigma model has a phase transition
for  $\beta < \beta_{c}\simeq 0.038$. 

In this paper we show that for any dimension $d>1$ the $H^{(2|2)}$ 
model exhibits localization 
for $\beta^{1/2}\ln \beta^{-1} \leq 1 /(2d-1)$. 
Thus the sigma model approximation
captures  the physics of both localization and diffusion.
Moreover, for a one dimensional chain we recover localization 
for all values of  
$\beta $. Localization is also expected in 2D (see \cite{DSZ} section 1.4 
and 4.3),   for all values of $\beta$ by both the  renormalization group 
and by a simple saddle analysis. 
However, a rigorous proof is still missing for this case. 

The techniques employed in this work to prove localization are quite 
different from the ones used in \cite{DSZ} to prove extended
states. The two papers can be read independently. The only
common point is the use of supersymmetry to prove some identities.
In the present case supersymmetry is applied only to prove that
the partition function is normalized to 1. 
We refer to  Section 4 and Appendix C in \cite{DSZ} for an introduction to
supersymmetric Ward identities.

\subsection{The model}

Let $\Lambda$ be a finite subset in $ \mathbb{Z}^{d}$ and $t_{j}$
a real variable for each site $j\in \Lambda $. We will consider periodic
or Neumann boundary conditions. Accordingly for any two points 
$x,y\in \lambda $,  $|x-y|$ will denote the Euclidian distance
on the lattice (Neumann bc) or the periodized lattice (periodic bc).

We introduce the probability measure
\begin{equation}\label{eq:tmeasure}
 d\mu_{\Lambda }^{\varepsilon} (t) = 
\ \prod_{j\in \Lambda} \frac{dt_{j}}{\sqrt{2\pi }} 
e^{-F_{\Lambda } (\nabla t)}e^{- M^{\varepsilon }_{\Lambda } (t)} 
\times 
\left[\sqrt {\det D^{\varepsilon }_{\Lambda } (t)}\   \right]
\end{equation}
where $dt_{j}$ is the Lebesgue measure, $F$ is the kinetic part and 
$M$ is the mass term:
\begin{align}
F_{\Lambda } (\nabla t)  &= \beta \sum_{(jj')\in \Lambda }  
(\cosh(t_j - t_{j'})-1) \cr
M^{\varepsilon }_{\Lambda } (t) &=  
\sum_{j\in \Lambda }  \varepsilon_{j} (\cosh t_j -1)\ . 
\label{eq:FG}
\end{align}
We denoted by $(jj')$ the nearest neighbor pairs $|j-j'|=1$. 
 $D^{\varepsilon }_{\Lambda } (t)$ is a positive definite matrix  defined by
\begin{equation}\label{eq:Mmatrix}
\begin{array}{lll}
(D^{\varepsilon }_{\Lambda })_{ij} & = 0  & |i-j|>1\\
(D^{\varepsilon }_{\Lambda })_{ij}   & = -\beta   & |i-j|=1\\
(D^{\varepsilon }_{\Lambda })_{jj}  & = \beta\left[2d+V_{j}\right]
+ \varepsilon_{j} e^{-t_{j}} & i=j\ ,\\
\end{array}
\end{equation}
\begin{equation}\label{eq:Vdef}
V_{j} =  \sum_{k, (jk)}  \left[e^{t_{k}-t_{j}} -1 \right]
\end{equation}
and $\varepsilon_{j}\geq 0$ are regularizing parameters
that are necessary to make the integral well defined.
We remark that $V_{j}+2d>0$ for all $t$ configuration and all
$j$.
Finally $\beta >0$ is a parameter that can be interpreted
as a measure of the temperature or the disorder. 

Note that the definition of the matrix $D$ differs from the
one introduced in \cite{DSZ} eq. (1.1). It is not difficult
to see that when you mix the term $\sum_{k}t_{k}$ and the determinant
in the effective
action (1.2) of \cite{DSZ} the result
is indeed the determinant of the  matrix $D$ above, more precisely
\begin{equation}\label{eq:Adef}
A_{ij} = e^{t_{i}} D_{ij} e^{t_{j}} = 
\left\{\begin{array}{ll}
0 &\ |i-j|>1\\
- \beta e^{t_{i}+t_{j}} & \ |i-j|=1\\
\beta \sum_{i', (ii')} e^{t_{i}+t_{i'}} & \ i=j\\
\end{array} \right.
\end{equation}
where $A$ is the matrix  introduced in \cite{DSZ} eq. (1.1).
For technical reasons the above representation is more convenient
when we want to prove diffusion as in \cite{DSZ} while the other
is more practical when we study localization (except in the proof  
of Theorem~2 where we will go back to the ``diffusion'' representation
for a while).
\medskip

Now, for any function $f (t)$ we will define its average by
\begin{equation}\label{eq:av}
\langle f (t) \rangle \ = \ \int  d\mu^{\varepsilon }_{\Lambda } (t) \ f (t)\ .
\end{equation}

\paragraph{Normalization  and choice of $\varepsilon_{j}$} 
By internal supersymmetry (see \cite{DSZ} Sect. 4 and eq.(5.1)) this
measure is already normalized to 1 so the partition function is
\begin{equation}\label{eq:Z}
Z^{\varepsilon }_{\Lambda } = \int\  d\mu^{\varepsilon }_{\Lambda } (t) \  = 1
\ .
\end{equation}
This identity is true regardless of the boundary conditions
and the values of $\beta $ or $\varepsilon_{j}$ as long as
the integral is well defined. Since we consider $\beta>0$ fixed
we only need $\varepsilon_{j}$ to be non zero at \textit{one lattice point}.
In the following we will consider three cases.
\begin{enumerate}
\item{} Uniform pinning: $\varepsilon_{j}= \varepsilon \leq 
\frac{1}{|\Lambda |}$ for all $j\in \Lambda $. The measure is
translation invariant with periodic bc. The correlation
function in this case has a divergent prefactor $1/\varepsilon$ in the
localized regime. 

\item{} Two pinnings: $\varepsilon_{x}=\varepsilon_{y}=O (1)$
and $\varepsilon_{j}=0$ for all other points. This is the analog of
inserting two electrical contacts in a metal sample. 

\item{} One pinning at $j=0$:  $\varepsilon_{0}=O (1)$ and 
$\varepsilon_{j}=0$ for all other points.  This is more suitable for 
an interpretation of the model as a random walk in a random environment.
Our results suggest that edge reinforced random walk (see
\cite{MR}) will also localize when the reinforcement is strong.

\end{enumerate}

\paragraph{The observable.} We will study the correlation function
\begin{equation}\label{eq-G}
G_{xy} := D^{-1}_{xy}
\end{equation}
where $x,y$ can be
any two points on the lattice such that both 
$\varepsilon_{x}>0 $ and $\varepsilon_{y}>0$. This observable
does not give information on localization properties in the
case of one pinning point. In such case a good observable
to study is
\begin{equation}\label{eq:obs1p}
{\cal O }_{j} = e^{+t_{j}/2}
\end{equation}
where $j$ is any point in the lattice.
This observable is analogous to  $x^{1/4}_{e} $ in the notation
of \cite{MR} where $e$ is an an edge $(j,j')$.


\subsection{Main results}

\begin{theor}\label{th-1}
Let $\varepsilon_{x}>0$, $\varepsilon_{y}>0$
 and $\sum_{j\in \Lambda }\varepsilon_{j}\leq 1$. 
Then for all $0<\beta <\beta_{c}$  ($\beta_{c} $  defined
below)   the correlation
function $G_{xy}$ \eqref{eq-G} decays exponentially 
with the distance $|x-y|$. More precisely: 
\begin{equation}\label{eq:decay1}
\langle G_{xy}\rangle \leq  \
C_{0} \left( \varepsilon_{x}^{-1} + \varepsilon_{y}^{-1}\right) \,
\left[I_{\beta}\, e^{\beta (c_{d}-1)}\, c_{d}  \right]^{|x-y|}\ ,
\end{equation}
where $c_{d}=2d-1$,    
$C_{0}$ is a constant and
\begin{equation}\label{xidef}
I_{\beta} = \sqrt{\beta } \int_{-\infty}^{\infty} \frac{dt}{\sqrt{2\pi } }
 e^{-\beta  (\cosh t -1) } \ .
\end{equation}
Finally  $\beta_{c}$ is defined by:
\begin{equation}\label{eq:betac}
I_{\beta_{c}} e^{\beta_{c} (c_{d}-1)} c_{d}  =1 \ .
\end{equation}
Our estimates hold uniformly in the volume.
\end{theor}

\paragraph{Remark 1} 
The integral $I_{\beta }$ satisfies $I_{\beta }<1$ $\forall \beta >0$:
\begin{equation}\label{eq:Ib}
I_{\beta } = \sqrt{\tfrac{\beta }{2\pi }} 
\int_{-\infty}^{\infty} dt\,
 e^{-\beta  (\cosh t -1) }\ <\ 
 \sqrt{\tfrac{\beta }{2\pi }}  \int_{-\infty}^{\infty} dt\, 
\cosh (t/2)\, e^{-\beta  (\cosh t -1) } \ = \ 1\ . 
\end{equation}
More precisely
\begin{equation}\label{eq:Iest}
I_{\beta } \leq \  
\left\{\begin{array}{ll}
 (\ln \beta^{-1} )\sqrt{\beta } & \beta <0.15\\
c e^{-\frac{1}{\beta }} & \beta >>1\\
\end{array}
 \right. 
\end{equation} 
where $c>1$ is a constant. 

\paragraph{Remark 2}
The constraints $\varepsilon_{x}>0$ and $\varepsilon_{y}>0$  exclude the case 
of one pinning. Moreover $\sum_{j\in \Lambda }\varepsilon_{j}\leq 1$
implies $\varepsilon\leq \frac{1}{|\Lambda |} $ when 
$\epsilon_{j}=\varepsilon $ is constant. The case of one pinning 
is covered by Theorem~2 below.

\paragraph{Main consequence}
For $d=1$ the critical beta is $\beta_{c}=\infty$ since 
\begin{equation}\label{eq:betac1}
I_{\beta } e^{\beta (c_{d}-1) } c_{d} = I_{\beta }<1\quad \forall \beta>0 
\ .
\end{equation}
Therefore the correlation function decays exponentially for all 
values of $\beta $. 
On the other hand for $d>1$ we obtain localization only for small 
$\beta $ since $\beta_{c}< (2d-1)^{-2}<1$.

\begin{theor}\label{th-2}
Let $\varepsilon_{0}=O (1)$  and $\varepsilon_{j}=0$ $\forall j\neq 0$.
Then for all $0<\beta <\beta_{c}$  ($\beta_{c} $  defined
below), the field $t_{x}$ wants to be as negative as $-|x|$.
More precisely $\langle {\cal O}_{x}\rangle$  (defined in \eqref{eq:obs1p})
decays exponentially
with the distance $|x-y|$ 
\begin{equation}\label{eq:decay3}
\langle {\cal O}_{x}\rangle \leq  \  C_{0}
\,\left[I_{\beta}\, e^{\beta (c_{d}-1)}\, c_{d}  \right]^{|x|}  \ ,
\end{equation}
where $c_{d}$ and $I_{\beta }$ are defined in Theorem~1 above
and $C_{0}$ is a constant.  Finally 
 $\beta_{c}$ is defined by:
\begin{equation}\label{eq:betacc}
I_{\beta_{c}} e^{\beta_{c} (c_{d}-1)} c_{d} =1\ . 
\end{equation}
Our estimates hold uniformly in the volume.
\end{theor}

\paragraph{Main consequence.}
For $d=1$ the critical beta is $\beta_{c}=\infty$ since 
\begin{equation}
I_{\beta } e^{\beta (c_{d}-1)}  c_{d} = I_{\beta }<1\quad \forall \beta >0\ .
\end{equation}
Therefore  $\langle {\cal O}_{x}\rangle$  decays exponentially for all 
values of $\beta $. 
On the other hand for $d>1$ the result holds only for small 
$\beta $ since $\beta_{c}< (2d-1)^{-2}<1$.


\paragraph{Acknowledgments.} It is our pleasure to thank 
A.~Abdesselam for discussions and 
suggestions related to this paper. A special thanks to M.~Zirnbauer
who explained the model to us and shared his many insights.

\section{Proof of Theorem \ref{th-1}}
\resetequ

We want to estimate
\begin{equation}\label{eq:Gav}
\langle G_{xy}\rangle\ =\ \int d\mu^{\varepsilon }_{\Lambda } (t)   G_{xy}
=\int \prod_{j\in \Lambda} \frac{dt_{j}}{\sqrt{2\pi }} 
e^{-F_{\Lambda } (\nabla t)}e^{- M^{\varepsilon }_{\Lambda } (t)} 
\times 
\sqrt{\det D^{\varepsilon }_{\Lambda } (t)} \  D^{-1}_{xy}\ .
\end{equation}

The proof in done in four steps.


\paragraph{Step 1.} We mix the observable $G_{xy}$ and a piece of the
probability measure namely $\sqrt{ \det D}$. The key identity is
\begin{equation}\label{eq:ident}
\sqrt{\det D^{\varepsilon }_{\Lambda } (t)} \  D^{-1}_{xy}=
\sqrt{ D^{-1}_{xy} }\ \sqrt{ D^{-1}_{xy}\det D^{\varepsilon }_{\Lambda } (t) }
\end{equation}
(remember that $D_{xy}^{-1}>0$).
The first term is bounded by
\begin{equation}\label{eq:Gxyb}
D^{-1}_{xy} \leq
\left( \frac{1}{\varepsilon_{x} e^{-t_{x} }} +  
\frac{1}{\varepsilon_{y} e^{-t_{y} }} \right)  \ .
\end{equation} 
This  is proved in Lemma \ref{lemma1}.
Inserting this in \eqref{eq:Gav} we have 
\begin{equation}\label{eq:1}
\langle G_{xy}\rangle\ \leq \ 
 \int d\nu^{\varepsilon }_{\Lambda } (t)  
\sqrt {[  D^{-1}_{xy}  \det D^{\varepsilon }_{\Lambda } (t)]} 
\left( \frac{e^{t_{x}/2}}{\sqrt{\varepsilon_{x} }}
+ \frac{e^{t_{y}/2}}{\sqrt{\varepsilon_{y} } }  \right)
\end{equation}
where 
\begin{equation}\label{eq:numeasure}
 d\nu^{\varepsilon }_{\Lambda } (t)  =  \prod_{j\in \Lambda}
 \frac{dt_{j}}{\sqrt{2\pi }}  
e^{-F_{\Lambda } (\nabla t)}e^{- M^{\varepsilon }_{\Lambda } (t)} \ .
\end{equation}
Unlike the measure $d\mu^{\varepsilon }_{\Lambda}$ given by 
\eqref{eq:tmeasure}, this measure is no longer normalized to 1.


\paragraph{Step 2.} We need to extract some decay in the distance $|x-y|$. 
This is hidden in $D^{-1}_{xy}\,{\det} D$. By some combinatorial arguments
(the proof is given in Lemma \ref{lemma2} )
we can write 
\begin{equation}\label{eq:2}
[ D^{-1}_{xy}  {\det} D^{\varepsilon }_{\Lambda } (t)]\ 
=\  \sum_{\gamma_{xy}}
\beta^{|\gamma |}  \ 
\det D^{\tilde{\varepsilon }}_{\Lambda^{c}_{\gamma }   }
\end{equation}
where the sum ranges over  non self intersecting paths $\gamma$ made of nearest
neighbor pairs in $\Lambda $ starting at $x$ and ending at $y$.  
Let  $|\gamma|,$ denote the length of $\gamma$ and let
$\Lambda _{\gamma } $ be the corresponding set of lattice points
and set $\Lambda _{\gamma }^{c} =\Lambda\backslash \Lambda _{\gamma }$. 
Finally $D^{\tilde{\varepsilon }}_{\Lambda^{c}_{\gamma }  }$  is the matrix
one obtains by deleting the rows and columns corresponding
to the lattice points $j\in \Lambda_{\gamma }$. It is 
is exactly like the matrix
$D^{\varepsilon }_{\Lambda }$, but defined on the complement of $\gamma $,
$\Lambda _{\gamma }^{c}$ and with modified masses:
\begin{equation}
\begin{array}{lll}
D^{\tilde{\varepsilon }}_{ij} & = 0  & |i-j|>1\\
D^{\tilde{\varepsilon }}_{ij}  & = -\beta   & |i-j|=1\\
D^{\tilde{\varepsilon }}_{ii}  & = \beta\left[ 2d + \tilde{V}_{i}
 \right]
+ \tilde{\varepsilon }_{i} e^{t_{i}} & i=j\ ,\\
\end{array}
\label{eq:Dtilde}\end{equation}
where $i,j\in \Lambda^{c}_{\gamma} $ and 
\begin{align}
\tilde{V}_{i} &=  \sum_{k\in\Lambda^{c}_{\gamma} , (ki)}
\left[ e^{t_{k}-t_{i}} -1\right] \\
 \tilde{\varepsilon }_{i}&=  \varepsilon_{i} + \beta 
\sum_{k\in \Lambda_{\gamma }, (ki)} e^{t_{k}}\ .\label{eq:Vtilde}
\end{align}
By combining \eqref{eq:1}  and \eqref{eq:2}   we have
\begin{equation}\label{eq:3}
\langle G_{xy}\rangle\ \leq \  
 \sum_{\gamma_{xy}} \beta^{|\gamma |/2 }     
\int d\nu^{\varepsilon }_{\Lambda } (t)  
\left( \frac{e^{t_{x}/2}}{\sqrt{\varepsilon_{x} }}
+ \frac{e^{t_{y}/2}}{\sqrt{\varepsilon_{y} } }  \right)
\sqrt {\det D^{\tilde{\varepsilon }}_{\Lambda^{c}_{\gamma }}} \ .
\end{equation}


\paragraph{Step 3.}  The measure $ d\nu^{\varepsilon }_{\Lambda } (t) $
defined in \eqref{eq:numeasure}  can be factored as a measure on 
$\Lambda_{\gamma } $ 
times a measure on the complement set $\Lambda_{\gamma }^{c} $
\begin{equation}
 d\nu^{\varepsilon }_{\Lambda } (t)\ = \
 d\nu^{\varepsilon }_{\Lambda_{\gamma } } (t) 
\ d\nu^{\varepsilon }_{\Lambda_{\gamma }^{c} } (t) \ 
 e^{-F_{\partial\gamma  } (\nabla t)} \ ,
\end{equation}
where 
\begin{equation}\label{eq:Fborder}
F_{\partial\gamma  } (\nabla t)\  =\  
 \sum_{(j,k), k\in \Lambda_{\gamma }, j\not\in \Lambda_{\gamma }} 
\beta\, (\cosh (t_{j}-t_{k})-1)
\end{equation}
describes the interaction between $\Lambda_{\gamma } $ and 
$\Lambda_{\gamma }^{c} $. Then the integral in  \eqref{eq:3}
can be written as
\begin{equation}\label{eq:4}
\int d\nu^{\varepsilon }_{\Lambda } (t)  
\left( \frac{e^{t_{x}/2}}{\sqrt{\varepsilon_{x} }}
+ \frac{e^{t_{y}/2}}{\sqrt{\varepsilon_{y} } }  \right)
\sqrt {\det D^{\tilde{\varepsilon }}_{\Lambda^{c}_{\gamma }}} \ = \ 
 \int d\nu^{\varepsilon }_{\Lambda_{\gamma } } (t)  \ 
\left( \frac{e^{t_{x}/2}}{\sqrt{\varepsilon_{x} }}
+ \frac{e^{t_{y}/2}}{\sqrt{\varepsilon_{y} } }  \right)
\  Z^{\gamma }_{\Lambda^{c}_{\gamma }} (t_{\gamma })\ ,\end{equation}
where we defined
\begin{align}\label{eq:Zcond}
Z^{\gamma }_{\Lambda^{c}_{\gamma }}(t_{\gamma }) &=\ 
\int d\nu^{\varepsilon }_{\Lambda^{c}_{\gamma } } (t)\ 
\sqrt {\det D^{\tilde{\varepsilon }}_{\Lambda^{c}_{\gamma }}}\ \ 
  e^{-F_{\partial\gamma  } (\nabla t)}\nonumber\\
&= \int \prod_{j\in\Lambda^{c}_{\gamma }}\  \frac{dt_{j}}{\sqrt{2\pi }} 
\  e^{-F_{\Lambda^{c}_{\gamma } } (\nabla t)}\  
e^{- M^{\varepsilon }_{\Lambda^{c}_{\gamma } } (t)}\  
 \sqrt{ {\det} D^{\tilde{\varepsilon }}_{\Lambda^{c}_{\gamma }  } }
\ \   e^{-F_{\partial\gamma  } (\nabla t)} \ .
\end{align}
Note that $Z^{\gamma }_{\Lambda^{c}_{\gamma }}(t_{\gamma })$ is 
still a function of the $t$ variables along the path 
$\{t_{k} \}_{k\in \Lambda_{\gamma } }$ (they are not integrated).
Now $Z^{\gamma }_{\Lambda^{c}_{\gamma }}(t_{\gamma })$
   is almost equal to the partition function 
\begin{align}\label{eq:Zcond1}
1= Z^{\tilde{\varepsilon }}_{\Lambda^{c}_{\gamma }} =&
 \int 
d\mu^{\tilde{\varepsilon } }_{\Lambda^{c}_{\gamma } }  (t)  =  
 \int d\nu^{\tilde{\varepsilon }}_{\Lambda^{c}_{\gamma } } (t)
\sqrt {\det D^{\tilde{\varepsilon }}_{\Lambda^{c}_{\gamma }}} \nonumber \\
=& \int \prod_{j\in\Lambda^{c}_{\gamma }} \frac{dt_{j}}{\sqrt{2\pi }} 
\  e^{-F_{\Lambda^{c}_{\gamma } } (\nabla t)} 
e^{- M^{\tilde{\varepsilon }}_{\Lambda^{c}_{\gamma } } (t)} \
 \sqrt{ {\det} D^{\tilde{\varepsilon }}_{\Lambda^{c}_{\gamma }  } }\ ,
\end{align}
where
$Z^{\tilde{\varepsilon }}_{\Lambda_{\gamma }^{c}} =1$  by supersymmetry 
(see \eqref{eq:Z}). 
Comparing  \eqref{eq:Zcond} and \eqref{eq:Zcond1}  
we see that there are two main differences:
\begin{enumerate}
\item [-] the mass term $M^{\varepsilon}_{\Lambda^{c}_{\gamma } }$
in  \eqref{eq:Zcond} depends on $\varepsilon $ instead of 
$\tilde{\varepsilon }$ and is smaller than what it should be
\[
M^{\varepsilon}_{\Lambda^{c}_{\gamma } }\leq 
M^{\tilde{\varepsilon }}_{\Lambda^{c}_{\gamma } } 
\]
since $M^{\tilde{\varepsilon }}$ contains additional mass terms;
\item [-] the exponent in \eqref{eq:Zcond} contains the additional factor  
 $-F_{\partial\gamma  } (\nabla t)$ coming from the kinetic interaction
between points on $\Lambda_{\gamma }$ and points on $ \Lambda^{c}_{\gamma }$.
\end{enumerate}
This last term is helping us since it makes the integral
smaller. We will use it to recover the missing mass. 
This is done in Lemma \ref{lemma3} below. The key ingredient
is a global translation on the $t$ variables.
The result is 
\begin{equation}\label{eq:5}
\langle G_{xy}\rangle\ \leq \  e^{1}\ 
 \sum_{\gamma_{xy}}    \beta^{|\gamma |/2}  
 e^{\beta |\partial\gamma | }
  \int d\nu^{\varepsilon }_{\Lambda_{\gamma } } (t)
  \left( \frac{e^{t_{x}/2}}{\sqrt{\varepsilon_{x} }}
+ \frac{e^{t_{y}/2}}{\sqrt{\varepsilon_{y} } }  \right)\ ,
\end{equation}
where $|\partial\gamma |\leq (2d-2)|\gamma |+2$ is the number of points
inside  $\Lambda_{\gamma }^{c} $ on the
boundary with $\Lambda_{\gamma }  $.


\paragraph{Step 4.} We are left with an integral along the path $\gamma $.  
The integral in \eqref{eq:5} is bounded by
\[
 \int d\nu^{\varepsilon }_{\Lambda_{\gamma } } (t) 
  \left( \frac{e^{t_{x}/2}}{\sqrt{\varepsilon_{x} }}
+ \frac{e^{t_{y}/2}}{\sqrt{\varepsilon_{y} } }  \right)
\ \leq\ \left(  \frac{I^{x}_{1}}{\sqrt{\varepsilon_{x} }}
 + \frac{I^{y}_{1}}{\sqrt{\varepsilon_{y} }}
 \right)  \ I_{2 }^{|\gamma|}\ ,
\]
where 
\begin{align}\label{eq:Iint}
I^{x}_{1} &= \int_{-\infty}^{\infty} \frac{dt}{\sqrt{2\pi } }
e^{t/2} e^{-\varepsilon_{x} (\cosh t -1) } = \frac{1}{\sqrt{\varepsilon_{x} }} 
\\
I_{2 } &= \int_{-\infty}^{\infty} \frac{dt}{\sqrt{2\pi } }
 e^{-\beta  (\cosh t -1) } = \frac{1}{\beta^{1/2} } I_{\beta }\ 
\end{align}
where $I_{\beta }$ was defined in \eqref{eq:Iint}. 
In the same way $I_{1}^{y}= 1/\sqrt{\varepsilon_{y} }$. 
Inserting all this we have
\begin{align}
\langle G_{xy}\rangle\ &\leq \  e^{1}
\left(\frac{1}{\varepsilon_{x} } + \frac{1}{\varepsilon_{y} } \right)
\sum_{\gamma_{xy}} \  e^{\beta |\partial\gamma | }\ 
(I_{\beta })^{|\gamma|}\label{eq:fb} \\
&\leq \ 2\ e^{1+2\beta }
\left(\frac{1}{\varepsilon_{x} } + \frac{1}{\varepsilon_{y} } 
\right) \sum_{n\geq |x-y|}   
\left( 
  e^{\beta (2d-2)} \, I_{\beta }  
\right)^{n} c_{d}^{n}\cr
& \leq  C_{0} \left(\frac{1}{\varepsilon_{x} } + \frac{1}{\varepsilon_{y} } 
\right) \,
\left[\, e^{\beta (2d-2) } I_{\beta }\, c_{d} 
\right ]^{|x-y|} \nonumber
\end{align}
where $c_{d}= (2d-1)$,  $C_{0}$ is a constant and the second 
inequality holds since the
number of self-avoiding walks made of $n$ steps is bounded by
$2d(2d-1)^{n}<2c_{d}^{n}$ and $|\partial \gamma |\leq (2d-2)|\gamma |+2$. 
Finally the sum over $n$ is convergent since
$(e^{\beta(2d-2)}I_{\beta}c_{d})~<~1$. 

This concludes the proof of Theorem 1.\qed 

\subsection{The lemmas}


\begin{lemma}\label{lemma1}
The following inequality holds:
\begin{equation}\label{eq:bound}
D^{-1}_{xy} \leq\   \frac{e^{t_{x}}}{\varepsilon_{x} } +
 \frac{e^{t_{y}}}{\varepsilon_{y} } \ .
\end{equation}
\end{lemma}


\paragraph{Proof} 
By Cauchy-Schwartz inequality
\[
D^{-1}_{xy}\ \leq \ 
\sqrt{D^{-1}_{xx} } \ \sqrt{D^{-1}_{yy} } \leq \ D^{-1}_{xx}  +  D^{-1}_{yy}
\ .
\]
Since $(f,Df)\geq \ \sum_{j} \varepsilon_{j} e^{-t_{j}} f_{j}^{2}$
for any $f_{j}\in \mathbb{R}$ we have
\[
D^{-1}_{xx} \leq \  \frac{1}{\varepsilon_{x}e^{-t_{x}} } \ .
\]
Hence the result.
 \qed


\begin{lemma}\label{lemma2}
For any invertible matrix $M$ on $\Lambda $ we have the following
identity:
\begin{equation}\label{eq:id}
[ M^{-1}_{xy}  {\det} M]\ 
=\  \sum_{\gamma_{xy}= (j_{1},\dotsc j_{m})} 
\left[ ( -M_{xj_{1}})  ( -M_{j_{1}j_{2}}) \cdots ( -M_{j_{m}y})  \right]\ 
{\det}_{\Lambda_{\gamma }^{c} } M   
\end{equation}
where  $\gamma $ is any non self intersecting path starting at $x$
and ending at $y$.
\end{lemma}


\paragraph{Proof} 
This is a classical formula arising from the fact that every permutation
can be decomposed as a product of cycles. One may derive it using the
representation of a determinant as a sum over a gas of disjoint non
self intersecting closed paths:
\[
\det M = \sum_{L_{1},\dotsc L_{p}} {\cal A} (L_{1})\cdots {\cal A} (L_{p})
\]
where  a loop $L= (j_{1},\dotsc ,j_{m})$  is an ordered set of
$m$ distinct points and 
\begin{equation}
{\cal A} (L) = 
\left\{\begin{array}{ll}
- \left[(-M_{j_{1}j_{2}}) (-M_{j_{2}j_{3}}) \cdots (-M_{j_{m}j_{1}}) \right]
& \ m>1\\
\quad M_{j_{1}j_{1}} &\  m=1\\
\end{array} \right.
\end{equation}
The sign $(-1)^{m-1}$ is the number of pairs inside the loop that 
need to be exchanged in order to recover the trivial permutation.
Now since $[ M^{-1}_{xy}  {\det} M]= \frac{\partial}{\partial M_{yx}} 
{\det} M $, the derivation selects only loops that contain the pair $yx$.
The corresponding matrix element disappears and the loop becomes
a path from $x$ to $y$. The sign $-1$ from $-M_{yx}$ cancels the global
$-1$ in front of the product. \qed 


\begin{lemma}\label{lemma3}
For any configuration of $\{t_{k}|\ k\in \Lambda_{\gamma } \}$, 
the conditioned partition function  
$Z^{\gamma }_{ \Lambda^{c}_{\gamma }} (t_{\gamma })$ given by \eqref{eq:Zcond} 
is bounded by
\begin{equation}\label{eq:Zbound}
 Z^{\gamma }_{ \Lambda^{c}_{\gamma }} (t_{\gamma }) \leq  
e^{\beta \sum_{k\in \Lambda_{\gamma }}  d^{\gamma }_{k} (1-e^{t_{k}-t^{*}}) }
e^{\sum_{j\in \Lambda_{\gamma }^{c} } \varepsilon_{j} (1-e^{-t^{*}}) }\leq\ 
e^{\beta|\partial\gamma | }
e^{\varepsilon |\Lambda^{c}_{\gamma }|}\ ,
\end{equation}
where $t^{*}$ is any real number satisfying
\begin{equation}\label{eq:tbound}
t^{*}\geq 0\ ,\qquad \mbox{and}\qquad t^{*}\geq t_{k}\qquad  \forall k\in 
\Lambda_{\gamma } \ ,
\end{equation}
and $ d^{\gamma }_{k}$ is the number of points nearest neighbor 
to $k$ that do not belong to $\Lambda_{\gamma }$:
\[
 d^{\gamma }_{k} = \#\{j\not\in \Lambda_{\gamma }\ |\  |j-k|=1\} \ .
\]
\end{lemma}


\paragraph{Proof}
Before doing any bound we perform a global translation inside 
the integral: 
\begin{equation}
t_{j}\ \rightarrow\ t_{j}+ t^{*}\qquad \forall j\in 
\Lambda^{c}_{\gamma }\ .
\end{equation}
Then inside the exponential we have:
\begin{align}
F_{\Lambda^{c}_{\gamma } } (\nabla t)  & \mapsto 
F_{\Lambda^{c}_{\gamma } } (\nabla t)\\
M^{\varepsilon }_{\Lambda^{c}_{\gamma } } (t) 
 & \mapsto M^{\varepsilon }_{\Lambda^{c}_{\gamma } } (t+t^{*}) \\
F_{\partial\gamma  } (\nabla t)& \mapsto 
F_{\partial\gamma  } (\nabla t+t^{*})=\hspace{-0.5cm}
 \sum_{(j,k), k\in \Lambda _{\gamma }, j\in \Lambda^{c}_{\gamma }} 
\hspace{-0.5cm}\beta\, (\cosh (t_{j}+t^{*}-t_{k})-1)\ .
\end{align}
In the numerator we have
\[
\left[ 
 \det D^{\tilde{\varepsilon }}_{\Lambda^{c}_{\gamma }  } 
\right] \mapsto 
\left[ 
 \det D^{\tilde{\varepsilon } e^{-t^{*}}}_{\Lambda^{c}_{\gamma }  } 
\right]
\]
so the only effect of the translation is to modify the mass 
$\tilde{\varepsilon }_{j}$ defined in \eqref{eq:Vtilde} at each lattice point. 
After the translation 
\begin{align}
&\hspace{-0,2cm}  Z^{\gamma }_{ \Lambda^{c}_{\gamma }}=
\int\left[ \prod_{j\in\Lambda^{c}_{\gamma }} \tfrac{dt_{j}}{\sqrt{2\pi }} 
 \right]
\  e^{-F_{\Lambda^{c}_{\gamma } } (\nabla t)} 
 \sqrt{ 
\det D^{\tilde{\varepsilon }e^{-t^{*}}}_{\Lambda^{c}_{\gamma }  } } 
e^{- M^{\varepsilon }_{\Lambda^{c}_{\gamma } } (t+t^{*})}
e^{-F_{\partial\gamma  } (\nabla t+t^{*})
}\label{eq:2.28}\\
& = \int \left[ \prod_{j\in\Lambda^{c}_{\gamma }} \tfrac{dt_{j}}{\sqrt{2\pi }} 
 \right]
 e^{-F_{\Lambda^{c}_{\gamma } } (\nabla t)} 
e^{- M^{\tilde{\varepsilon }e^{-t^{*}} }_{\Lambda^{c}_{\gamma }}
\hspace{-0.2cm} (t)}
  \sqrt{ 
\det D^{\tilde{\varepsilon }e^{-t^{*}}}_{\Lambda^{c}_{\gamma }  } } \ 
 e^{M^{\tilde{\varepsilon }e^{-t^{*}} }_{\Lambda^{c}_{\gamma }}
\hspace{-0.2cm} (t)
 - M^{\varepsilon }_{\Lambda^{c}_{\gamma } } (t+t^{*}) 
-F_{\partial\gamma  } (\nabla t+t^{*})}
\cr
& = \int d\mu^{ \tilde{\varepsilon }e^{-t^{*}} }_{\Lambda_{\gamma }^{c} }
\hspace{-0.2cm} (t)\ 
\   e^{\sum_{j\in\Lambda^{c}_{\gamma }  } {\cal E}rr_{j} } \ 
\leq\    Z^{\tilde{\varepsilon }e^{-t^{*}} }_{ \Lambda^{c}_{\gamma }}
\ e^{\sum_{j\in\Lambda^{c}_{\gamma }  }
\sup_{t_{j}} {\cal E}rr_{j}  } 
\ = \ \prod_{j\in\Lambda^{c}_{\gamma } } 
   e^{ \sup_{t_{j}} {\cal E}rr_{j}  } \ ,
\nonumber\end{align}
where  we used   
$ Z^{\tilde{\varepsilon }e^{-t^{*}} }_{ \Lambda^{c}_{\gamma }}=1$ 
(see \eqref{eq:Z}) and  we defined
\begin{equation}
\sum_{j\in\Lambda^{c}_{\gamma }  } {\cal E}rr_{j} = \left[
\ M^{\tilde{\varepsilon }e^{-t^{*}} }_{\Lambda^{c}_{\gamma } } (t)
- M^{\varepsilon }_{\Lambda^{c}_{\gamma } } (t+t^{*})\ 
- F_{\partial\gamma  } (\nabla t+t^{*})
 \right]\end{equation}
\begin{equation}
 {\cal E}rr_{j}  =
\tilde{\varepsilon}_{j} e^{-t^{*}} (\cosh t_{j}-1)\  -\ 
\varepsilon_{j}   (\cosh ( t_{j}+t^{*}) -1) 
\ -\   \hspace{-0.4cm} \sum_{k\in \Lambda_{\gamma }, (j,k)}
\hspace{-0.4cm}\beta\  (\cosh (t_{j}+t^{*}-t_{k})-1).
\end{equation}
To conclude we shall prove that the right hand side of \eqref{eq:2.28}  
is bounded by the rhs of \eqref{eq:Zbound}:
\[
 \ \prod_{j\in\Lambda^{c}_{\gamma } }   e^{ \sup_{t_{j}} {\cal E}rr_{j}  } 
\ \leq \ 
e^{\beta \sum_{k\in \Lambda_{\gamma }}  d^{\gamma }_{k} (1-e^{t_{k}-t^{*}}) }
e^{\sum_{j\in \Lambda_{\gamma }^{c} } \varepsilon_{j} (1-e^{-t^{*}}) }\ .
\]
We distinguish two cases.

\paragraph{Case 1.}
When $j$ is far from the path $\gamma $ that is $|j-k|>1$ for all
$k\in \Lambda_{\gamma }$ then $\tilde{\varepsilon}_{j}=\varepsilon_{j} $ 
and we have
\begin{align}
 {\cal E}rr_{j}&= \varepsilon_{j} \left[
e^{-t^{*}} (\cosh t_{j}-1) -  (\cosh ( t_{j}+t^{*}) -1) 
 \right]\cr
&
 = \varepsilon_{j} \left[ e^{t_{j}} \sinh (-  t^{*}) + 1 -  
e^{-t^{*}}  \right] \ \leq \ 
\varepsilon_{j} ( 1 -  e^{-t^{*}} ) \ ,
\end{align}
where the last inequality holds for $t^{*}\geq 0$.

\paragraph{Case 2.}
When $j$ is near to the path $\gamma $ that is $|j-k|=1$ for some
$k\in \Lambda_{\gamma }$ then $\tilde{\varepsilon}_{j}=\varepsilon_{j} +
 \sum_{k\in \Lambda_{\gamma }, (j,k)} \beta e^{t_{k}} $ and
we have
\begin{align}
 {\cal E}rr_{j} =&\varepsilon_{j} \left[
e^{-t^{*}} (\cosh t_{j}-1) -  (\cosh ( t_{j}+t^{*}) -1) 
 \right]  \\
&\quad + \beta \hspace{-0.5cm}
\sum_{k\in \Lambda_{\gamma }, (j,k)} 
\left[  e^{t_{k}-t^{*}}  (\cosh t_{j}-1)   -
 (\cosh (t_{j}+t^{*}-t_{k})-1)
 \right]\cr
&= \varepsilon_{j} \left[ e^{t_{j}} \sinh (-  t^{*}) + 1 -  
e^{-t^{*}}  \right] +  \beta 
\hspace{-0.5cm}\sum_{k\in \Lambda_{\gamma }, (j,k)}
 \left[ e^{t_{j}} \sinh (t_{k}-  t^{*}) + 1 -  e^{t_{k}-t^{*}}  \right]
\cr
&\leq   \varepsilon_{j} (1 - e^{-t^{*}}) + \beta 
\hspace{-0.5cm}\sum_{k\in \Lambda_{\gamma }, (j,k)}
( 1 -  e^{t_{k}-t^{*}})\ ,\nonumber
\end{align}
where the last inequality holds if  $t^{*}\geq 0$ and
$t^{*}\geq t_{k}$ $\forall k\in \Lambda_{\gamma }$.
\medskip

Finally
\begin{align}
\hspace{-1cm}\sum_{j\in\Lambda^{c}_{\gamma }  } {\cal E}rr_{j} 
&\leq 
 \ \sum_{j\in \Lambda^{c}_{\gamma }}\
[ \  \varepsilon_{j} (1 - e^{-t^{*}}) + \ \beta 
\hspace{-0.2cm}\sum_{k\in \Lambda_{\gamma }, (j,k)}
( 1 -  e^{t_{k}-t^{*}}) \  ]\cr
& \leq  \beta \sum_{k\in \Lambda_{\gamma }}  
d^{\gamma }_{k} (1-e^{t_{k}-t^{*}})
\ +\  \sum_{j\in \Lambda^{c}_{\gamma } }   \varepsilon_{j} (1-e^{-t^{*}})\ .
\end{align}
This concludes the proof of the lemma. \qed

\section{Proof of Theorem 2}
\resetequ 

The proof of Theorem~2 is almost identical to that of
Theorem~1.
This time there is no term $D^{-1}_{xy}$ ensuring we
can extract a path $\gamma $ connecting $x$ to $y$.
On the other hand, 
since there is a pinning only at one position the matrix-tree
theorem (see \cite{abdesselam} for a simple proof and many references) 
applied to the  ``diffusion'' representation $A$ given in \eqref{eq:Adef}  
 of the matrix $D$ gives 
\begin{equation}\label{eq:detp}
 {\det} A = \varepsilon_{0} e^{t_{0}}    \sum_{T} \prod_{jj'\in T}\,
(-A_{jj'}) = 
\varepsilon_{0} e^{t_{0}} \sum_{T}  
\prod_{(jj')\in T}\, \left[\beta e^{t_{j}+t_{j'}} \right]\ ,
\end{equation}
where the sum is over the spanning trees on $\Lambda $ made of 
nearest neighbor pairs (since $A_{jj'}=0$ when $|j-j'|>1$).
Therefore each term in the sum contains a path $\gamma $ from $0$ to $x$.
Actually using \eqref{eq:detp} it is easy to see that 
\begin{equation}
A^{-1}_{0x}\  {\det} A = \frac{1}{\varepsilon_{0}e^{t_{0}} }\  {\det} A
\end{equation}
for all $x\in \Lambda $, $x\neq 0$. Therefore 
$\varepsilon_{0}e^{t_{0}}  A^{-1}_{0x}=1$ and
\begin{equation}
{\det} D = \varepsilon_{0}e^{t_{0}}\  A^{-1}_{0x} \ {\det} D =
 \varepsilon_{0} e^{-t_{x}}\   D^{-1}_{0x} {\det} D =
 \varepsilon_{0} e^{-t_{x}}\ \sum_{\gamma_{0x} } \beta^{|\gamma |} 
  {\det}_{\Lambda_{\gamma }^{c} } D\ ,
\end{equation}
where $  A^{-1}_{0x}=e^{-t_{0}} D^{-1}_{0x}e^{-t_{x}}$ and
 in the last term we applied Lemma \ref{lemma2}. 
Inserting this result in 
\eqref{eq:tmeasure} we have
\begin{align}
&\langle {\cal O}_{x}\rangle \ 
= \  \int  d\mu^{\varepsilon }_{\Lambda } (t)
\  {\cal O}_{x}\ = \ 
 \int d\nu^{\varepsilon }_{\Lambda }(t) 
\sqrt{\det D^{\varepsilon }_{\Lambda } }\  e^{t_{x}/2} \\ 
&\  = \sqrt{\varepsilon_{0} } 
 \int d\nu^{\varepsilon }_{\Lambda }(t)   e^{t_{x}/2} e^{-t_{x}/2}
\sqrt{  \sum_{\gamma_{0x} } \beta^{|\gamma |}\,  
 {\det} D^{\tilde{\varepsilon }}_{\Lambda^{c}_{\gamma } } } \cr 
&\ \leq  \ \sqrt{\varepsilon_{0} }  \sum_{\gamma_{0x}} \beta^{|\gamma |/2} 
\int d\nu^{\varepsilon }_{\Lambda_{\gamma } } (t)
  \ Z^{\gamma }_{\Lambda^{c}_{\gamma }} (t_{\gamma })\ 
\leq \  \sqrt{\varepsilon_{0} }  \sum_{\gamma_{0x}} \beta^{|\gamma |/2} 
\int d\nu^{\varepsilon }_{\Lambda_{\gamma } } (t)\   
e^{\beta |\partial \gamma |}\  
\cr
&\ \leq   \sum_{\gamma_{0x}}\  
 e^{\beta |\partial \gamma |}\  
I_{\varepsilon_{0} } \  I_{\beta }^{|\gamma |}
\ \leq \ C_{0}\   (c_{d} e^{\beta (c_{d}-1)} I_{\beta })^{|x|} \ ,
\nonumber\end{align}
where  $ d\nu^{\varepsilon }_{\Lambda }(t) $ was defined 
in \eqref{eq:numeasure},  $Z^{\gamma }_{\Lambda^{c}_{\gamma }} (t_{\gamma })$ 
in \eqref{eq:Zcond},  $I_{\beta }$ in \eqref{eq:Ib}
and the same definition holds for $I_{\varepsilon  }$.
In the  second line we used  $ {\det}_{\Lambda_{\gamma }^{c} } D = 
{\det} D^{\tilde{\varepsilon }}_{\Lambda^{c}_{\gamma } } $  (see 
\eqref{eq:Dtilde}). We used  Lemma~3 
eq. \eqref{eq:Zbound} to bound 
$Z^{\gamma }_{\Lambda^{c}_{\gamma }}(t_{\gamma })$.  
Finally  the last inequality holds since $ (c_{d}e^{c_{d}-1} I_{\beta })<1$. 
This concludes the proof of  Theorem~2.\qed


\bibliographystyle{plain}



\end{document}